\documentclass{article}
\usepackage{amssymb,amsmath}
\usepackage{icrctc07}

\title{Hybrid Performance of the Pierre Auger Observatory}
\shorttitle{Hybrid Performance of the Pierre Auger Observatory}

\authors{B.R. Dawson$^1$, for the Pierre Auger Collaboration$^{2}$}
\shortauthors{B.R. Dawson et al.}
\afiliations{$^1$Department of Physics, University of Adelaide, Adelaide 5005, Australia\\ $^2$Observatorio Pierre Auger, Av. San Mart\'in Norte 304, (5613) Malarg\"{u}e, Mendoza, Argentina}

\email{bruce.dawson@adelaide.edu.au}

\abstract{
A key feature of the Pierre Auger Observatory is its hybrid design, in
which ultra high energy cosmic rays are detected simultaneously by
fluorescence telescopes and a ground array.  The two techniques see
air showers in complementary ways, providing important cross-checks
and measurement redundancy.  Much of the hybrid capability stems from
the accurate geometrical reconstruction it achieves, with accuracy
better than either the ground array detectors or a single telescope could
achieve independently.  We have studied the geometrical and
longitudinal profile reconstructions of hybrid events.  We present the
results for the hybrid performance of the Observatory, including
trigger efficiency, energy and angular resolution, and the efficiency
of the event selection.
}

\begin{document}
\maketitle


\vspace{-10mm}
\section{Introduction}

\begin{figure}[!bt]
\begin{center}
\noindent
\includegraphics [width=0.40\textwidth]{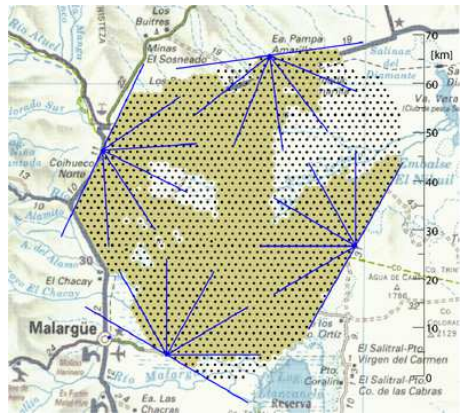}
\end{center}
\vspace{-5mm}
\caption{The Observatory in May 2007, showing the positions of the
four FD stations and the approximately 1200 deployed SD tanks (shaded
region).}\label{fig1}
\end{figure}

\vspace{-3mm} The Pierre Auger Observatory is located in the province
of Mendoza in western Argentina (35.5$^\circ$S, 69.3$^\circ$W).
Construction will be complete at the end of 2007, but production data
have been collected by the growing observatory since January 2004.  At
the time of writing, over 1200 of the 1600 water Cherenkov particle
detector tanks have been deployed on a 1.5\,km triangular grid
\cite{Suomijarvi} (Figure ~\ref{fig1}).  Each surface detector (SD)
tank contains 12 tonnes of water (10\,m$^2$ area), and each is
equipped with local digitizing electronics (400\,MHz sampling rate),
solar power, GPS receiver and a radio communication system \cite{EA}.
The final fluorescence detector (FD) site came into operation in
February 2007 on the northern edge of the SD array.  Now four sites
view the atmosphere above the array, with each site consisting of 6
Schmidt telescopes, a design chosen for improved optical performance.
The telescopes each have a field of view of approximately $30^\circ
\times 30^\circ$, mirror area of 12\,m$^2$, aperture area of
3.8\,m$^2$ and 440 hexagonal pixels of 1.5$^\circ$ diameter.  Pixel
signals are digitized with 100\,MHz sampling \cite{EA}.

The unique ``hybrid'' combination of fluorescence and surface
detectors has enormous advantages in all areas of the mission of the
Observatory \cite{hybadvantages}.  For example, in our studies of the
ultra-high energy cosmic ray (UHECR) energy spectrum \cite{Roth} the
SD provides the energy parameter S(1000), a huge collecting area,
24\,hr operation and an easily calculable aperture.  The FD provides
the conversion between S(1000) and the cosmic ray primary energy,
since the FD uses a near-calorimetric technique for determining
energy.  This avoids calibrating S(1000) via shower simulations, which
have uncertainties related to hadronic interaction models.  In
anisotropy studies, hybrid data provide high-precision shower arrival
directions which are used to cross-check SD-derived directions and to
directly measure the SD angular resolution.  In mass composition
studies, the FD measures the depth of shower maximum $X_\text{max}$,
the least indirect of all mass indicators \cite{Unger}.  Meanwhile,
hybrid data are being used to calibrate and cross-check several
promising mass sensitive parameters measured by the SD alone
\cite{Healy}.

The key to the success of hybrid observations is the precise
measurements of shower arrival directions.  Hybrid data supplements
the traditional FD direction fitting method with the arrival time of
the shower at the ground measured by a single SD tank.  Direction
resolution of better than $0.5^\circ$ not only makes it possible for
sensitive anisotropy searches and cross-checks of SD direction
assignments, it is also the first step towards high quality
measurements of shower longitudinal profiles, and the extraction of
$X_\text{max}$ and energy \cite{AstPartHybrid}.

\vspace{-5mm}
\section{Challenges of a Hybrid Observatory}

\vspace{-3mm} 
Some experimental challenges exist in fully realising the promise of
the hybrid technique in providing high quality measurements of shower
parameters.  Most are connected with the FD, since the only data taken
from the SD with this technique is the arrival time of the shower at a
single tank.  The challenges can be divided into three areas - those
related to the detector, those related to the atmosphere, and those
involved in the reconstruction procedure.

Detector-related challenges include the optical and electronics
calibration of the FD system, including its wavelength dependence.  We
employ an ``end-to-end'' technique which uses a uniformly illuminated drum
positioned at the entrance aperture of a telescope to provide the
conversion between a photon flux at the aperture and ADC counts in the
electronics \cite{Knapik}.  The drum is deployed periodically through
the year, and allows measurements at five wavelengths.  Nightly
relative measurements made with local fixed light sources keep track of any
changes between drum calibrations.  The current estimate of the
systematic uncertainty for shower energy related to the optical
calibration is 9.5\%.  The hybrid method also requires
calibration and monitoring of the telescope alignment, and the
synchronization of timing at FD sites and the SD tanks.  The former is
monitored with star positions and laser shots to a precision of
$0.05^\circ$; the latter is monitored and is known at a level
of approximately 100\,ns.

The atmosphere is our detection medium, and its properties must be
carefully monitored.  Fluorescence light is produced in proportion to
the energy deposited in the atmosphere by shower particles.  The
efficiency of light production has a dependence on pressure,
temperature and humidity.  Data from \cite{Nagano04} are currently
being applied, where the current systematic uncertainty in the
absolute fluorescence efficiency is 14\%, and an additional
uncertainty of 7\% is related to pressure, temperature and humidity
effects.  Improvements in these uncertainties are expected in the near
future.  The fluorescence light is emitted isotropically from the
excited molecules, and is attenuated on its way to the detector by
Rayleigh scattering off air molecules and by scattering due to
aerosols.  Average monthly models of the molecular atmosphere are
sufficient to take account of Rayleigh scattering, but treatment of
aerosol scattering requires hourly measurements of the characteristics
and distribution of aerosols \cite{BenZvi}.  The Observatory also uses
several techniques to detect night-time cloud.  The systematic
uncertainties in atmospheric attenuation contribute approximately 4\%
to the systematic uncertainty budget for hybrid estimates of shower
energy.

In the algorithm used to reconstruct the longitudinal profile of a
shower, one of the important steps is the collection of light in the
focal plane of the telescope.  Care must be taken to collect the
fluorescence light properly (including light from the full lateral
width of the shower) without risking the inclusion of night-sky light
that dominates away from the image axis.  Also, light received at the
detector includes direct and scattered Cherenkov light from the
atmosphere, which must be accounted for.  Systematic uncertainties in
these and other parts of the reconstruction method contribute 10\%
to the total uncertainty in the measured energy.  A final
correction to the energy takes account of the part of the
shower energy that does not contribute proportionally to fluorescence
light (e.g. neutrinos, high energy muons).  This energy-dependent and
mass-dependent ``invisible energy'' correction has a systematic
uncertainty of 4\% \cite{invis}.

Table 1 summarizes the systematic uncertainties in determining energy
by the hybrid method.

\begin{table}[t] 
\begin{center}
\begin{tabular}{|l|r|}
\hline 
Source & Systematic uncertainty \\
\hline
Fluorescence yield & 14\% \\
P,T and humidity  & 7\% \\
effects on yield &  \\
Calibration & 9.5\% \\
Atmosphere & 4\% \\
Reconstruction & 10\% \\
Invisible energy & 4\% \\
\hline 
TOTAL &  22\% \\
\hline 
\end{tabular} 
\caption{Systematic uncertainties in determining energy by the hybrid method.
Efforts are underway to reduce the main uncertainties in the fluorescence
yield, the absolute calibration, and in the reconstruction method.}
\label{table}
\end{center}
\end{table}

\vspace{-5mm}
\section{Trigger Efficiency and Event Selection}

\vspace{-3mm}
Hybrid triggers are formed in near real time, when triggers from
fluorescence telescopes are matched with local triggers from
individual SD tanks.  The local tank trigger, known as a ``T2'', 
is described in \cite{trigger}.

\begin{figure}[!bth]
\begin{center}
\noindent
\includegraphics [width=0.40\textwidth]{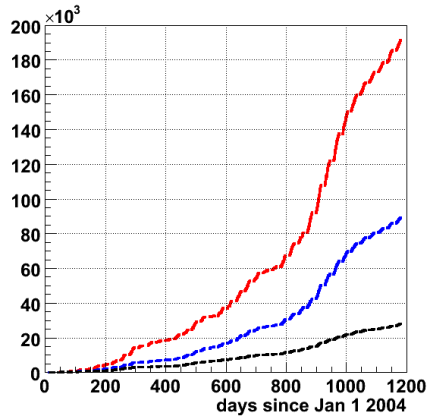}
\end{center}
\vspace{-5mm}
\caption{Growth of the hybrid data set since 2004. Shown are the
number of events with geometries successfully reconstructed (top
line), those where shower maximum $X_\text{max}$ is viewed (middle),
and those with reconstructed energies $>10^{18}$eV (bottom).}
\label{fig2}
\end{figure}

Simulations of the trigger efficiency have been performed, partly in
connection with a hybrid energy spectrum study \cite{Perrone}.  The
hybrid trigger is fully efficient across the entire SD array above
$10^{19}$eV, but a significant aperture is available down to energies
well below $10^{18}$eV.  Showers satisfying the triggering criteria can
generally be reconstructed to provide good arrival direction
information, but not all events provide good estimates of energy or
$X_\text{max}$.  For example, in the study of the energy dependence of
$X_\text{max}$ \cite{Unger}, cuts are required on the quality of the
observed longitudinal shower profile, as well as cuts to ensure that
showers in the sample were not biased in $X_\text{max}$ by the limited
range of elevations viewed by the FD telescopes.  The same ``quality
cuts'' were applied to showers used in the hybrid calibration of the
SD energy parameter S(1000) \cite{Roth}.

Figure~\ref{fig2} shows the growth of the hybrid data set since
January 2004, including all events successfully passing the geometry
reconstruction stage, and the number passing two other simple cuts.

\vspace{-5mm}
\section{Geometry and Profile Resolution}

\vspace{-3mm}
The line of triggered pixels in an FD camera defines a plane in space
containing the shower axis and a point representing the FD, known as
the shower-detector plane (SDP).  The orientation of the shower axis
within the SDP is determined using timing information.  With an FD
alone, the reconstruction of the axis within the SDP can sometimes
suffer from degeneracy related to the inability to detect changes in
the angular speed of the shower image across the FD camera.  The
hybrid technique breaks this degeneracy by including the arrival time
of the shower at ground level, data provided by a single SD tank near
the shower axis \cite{AstPartHybrid}.  

\begin{figure}[!bt]
\begin{center}
\noindent
\includegraphics [width=0.40\textwidth]{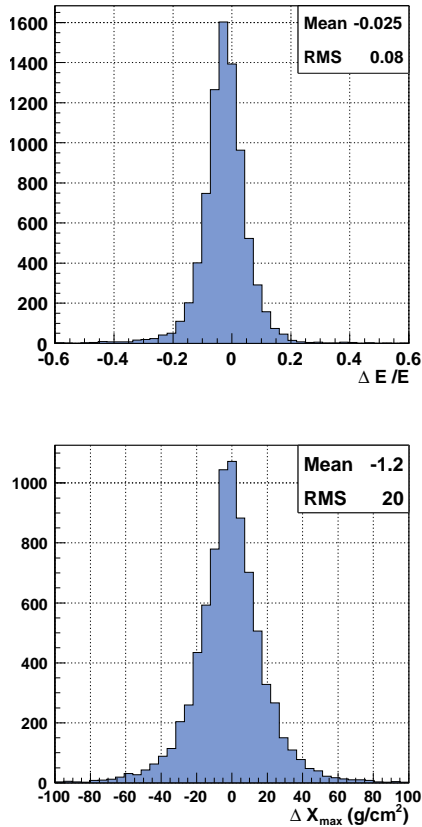}
\end{center}
\vspace{-5mm}
\caption{Resolution results for $10^{18}-10^{19}$eV ($E^{-2}$ differential
spectrum).  Applying cuts from \cite{Unger}, statistical resolution of 8\%
in energy and 20\,g\,cm$^{-2}$ in $X_\text{max}$ is achieved.}
\label{fig3}
\end{figure}

Simulations have been performed to estimate the geometry and shower
profile resolution.  A sample of showers with energies in the range
$10^{18}-10^{19}$eV have been simulated with a $E^{-2}$ differential
energy spectrum, thus including a rough allowance for the growing
FD aperture with energy.  With minimum cuts (angular track length
$>15^\circ$, reconstructed tank-core distance $<$ 2\,km) the median
and 90\% core location errors are 35\,m and 150\,m respectively, and
the median and 90\% arrival direction errors are $0.35^\circ$ and
$0.95^\circ$.  The profile resolution results are shown in
Figure~\ref{fig3}.  Quality cuts described in \cite{Unger} have been
applied for these plots.  

At the higher energies the observatory has measured a number of
showers observed by two (or more) FD sites.  This offers an
opportunity to cross-check these simulation results, though two
caveats apply.  First, the event statistics are low, especially after
standard quality cuts are applied to each of the views of a shower.
Secondly, the steeply falling energy spectrum means that many of these
``stereo'' events have a lower than average quality image in at least
one of the two FD eyes.  In any case, the single-eye energy and
$X_\text{max}$ resolution figures derived from stereo events (11\% and
18\,g\,cm$^{-2}$ respectively) are entirely consistent with simulation
results.

\end{document}